\documentstyle[preprint,prb,aps,epsf,eqsecnum]{revtex}
\begin{document}
%\draft
\tightenlines

\title{Functional Equations and Fusion Matrices for the Eight Vertex Model}

\author{Klaus Fabricius
\footnote{e-mail Fabricius@theorie.physik.uni-wuppertal.de}}
\address{ Physics Department, University of Wuppertal, 
42097 Wuppertal, Germany}
\author{Barry~M.~McCoy
\footnote{e-mail mccoy@insti.physics.sunysb.edu}}               
\address{ Institute for Theoretical Physics, State University of New York,
 Stony Brook,  NY 11794-3840}
\date{\today}
\preprint{YITPSB-03-59}

\maketitle

\begin{abstract}

We derive sets of functional equations for the eight vertex model by 
exploiting an analogy with the functional equations of the  chiral
Potts model. From these equations we show that the fusion matrices have special
reductions at certain roots of unity. We explicitly exhibit these reductions
for the 3,4 and 5 order fusion matrices and compare our formulation
with the algebra of Sklyanin.
 
\end{abstract}

\section{Introduction}

The solution of the 8 vertex model by Baxter \cite{bax1} has many inventive
steps. First of all the eigenvalues of the transfer matrix $T$ are studied
without having any information on the eigenvectors. Secondly these
eigenvalues are computed by inventing a new auxiliary matrix (called $Q$)
which commutes with $T$ and which satisfies a functional equation with
$T$. This functional equation has the further property that it can be
used to find all eigenvalues of $T$. The determination of these
eigenvalues is often what is called ``solving the 8 vertex model''

Many years after the eigenvalues of the 8 vertex model transfer matrix
 were computed a more elaborate model called the chiral Potts model was
discovered to be integrable \cite{ampty}. 
For this model there is also an auxiliary
matrix Q which satisfies a functional equation \cite{bs} with the transfer
matrix but unlike the 8 vertex model this functional equation is not
sufficient to compute the eigenvalues of the transfer matrix. 
However, further functional equations were first conjectured 
\cite{amp1},\cite{amp2} and
 then proven true \cite{bbp} from which the
desired eigenvalues are computed.

There are in fact many matrices $Q$ which satisfy the $TQ$ equations
first discovered \cite{bax1} in 1972 for the special ``root of unity'' case
\begin{equation}
\eta={m_1K\over L}
\label{root}
\end{equation}
 where the parameter $\eta$ of the 8 vertex model is a rational 
multiple of $K$ the  real period of the
elliptic functions. In 1973 Baxter \cite{bax2}-\cite{bax4} found a
second matrix $Q$ which satisfies the $TQ$ equations in the root of
unity case and also for the case of generic $\eta.$
We will here concentrate on the $Q$ matrix of the 1972 paper
\cite{bax1} which we denote by $Q_{72}.$

Recently we have discovered \cite{fm1} that in the root of unity case 
(\ref{root})
the functional equations discovered by
Baxter \cite{bax1} do not in fact 
exhaust the totality of functional equations of
the 8 vertex model and we conjectured a functional equation \cite{fm1} 
obeyed by $Q_{72}$ alone which allows the 
computation of the eigenvalues of $Q_{72}$ 
without reference  to $T$. When (\ref{root}) holds many of the 
eigenvalues of $T$
are degenerate and Baxter's $TQ$ functional equation is not sufficient to
compute the degeneracy of the eigenvalues. However the auxiliary matrix
$Q_{72}$ is non degenerate and the new functional equation of
ref. \cite{fm1} allows the
complete determination of the spectrum of $T$ including the degeneracies.

In the course of searching for a proof of the conjectured functional
equation of ref. \cite{fm1} it has become clear that the 
8 vertex model has several more
functional equations than were originally found in ref. \cite{bax1} and in
fact there is a complete analogy between the 8 vertex model and all
the functional equations found in ref. \cite{bbp} 
for the chiral Potts model. The
purpose of this paper is to present this analogy. 

The concept of the fusion hierarchy \cite{kure}-\cite{djkmo} 
will play an important role in
the analogy and will allow us to find a new functional equation for
the 8 vertex transfer matrix which has not yet appeared in the
literature. Curiously enough, for the chiral Potts model the
fusion matrices have been explicitly computed in ref. \cite{bbp} 
for all levels of fusion 
while for the
8 vertex model we are at present only
able to compute explicit formulas for the fusion levels 3,4 and 5.

In sec. 2 we review the derivation of the $TQ$ functional equation of
Baxter. In sec. 3 we present the analogy with the
chiral Potts model 
in detail and derive functional equations for the 8 vertex model. 
In sec. 4 we explicitly compute
the fusion matrices for levels 3,4 and 5.  
In sec. 5 we introduce a similarity transformation which makes explicit
the degeneration which takes place in $T^{(L+1)}(v)$ when $\eta$
satisfies the root of unity condition (\ref{root}).
In sec. 6 we compare our results with
the algebra of Sklyanin \cite{sk1},\cite{sk2}.
We close in sec. 7 with a proof of the
functional equation of ref.\cite{fm1} for the case of $L=2.$

\section{Baxter's $TQ$ equation}

We begin our considerations by reviewing Baxter's derivation of the
$TQ$  functional equation of ref.\cite{bax1}.
The transfer matrix for the eight vertex model with $N$ columns and periodic
boundary conditions is
\begin{equation}
T_8(u)|_{\bf \mu,\nu}={\rm Tr} W_8(\mu_1,\nu_1)W_8(\mu_2,\nu_2)
\cdots W_8(\mu_N,\nu_N)
\label{t8}
\end{equation}
where in the conventions of  (6.2) of ref. \cite{bax1}
\begin{eqnarray}
W_8(1,1)|_{1,1}=W_8(-1,-1)|_{-1,-1}&=&
\Theta(2\eta)\Theta(v-\eta)
H(v+\eta)=a(v)\nonumber\\
W_8(-1,-1)|_{1,1}=W_8(1,1)|_{-1,-1}&=&\Theta(2\eta)
H(v-\eta)\Theta(v+\eta)=b(v)\nonumber\\
W_8(-1,1)|_{1,-1}=W_8(1,-1)|_{-1,1}&=& H(2\eta)
\Theta(v-\eta)\Theta(v+\eta)=c(v)\nonumber\\
W_8(1,-1)|_{1,-1}=W_8(-1,1)|_{-1,1}&=& H(2\eta)
H(v-\eta)H(v+\eta)=d(v)
\label{bw}
\end{eqnarray}
The definition and useful properties of $H(v)$ and $\Theta(v)$ are
recalled in the appendix.

In ref.\cite{bax1} the matrix  $Q_R(v)$ is defined as
\begin{equation}
[Q_R(v)]_{\alpha |\beta}={\rm Tr}S_R(\alpha_1, \beta_1)
S_R(\alpha_2, \beta_2)\cdots S_R(\alpha_N, \beta_N)
\label{qrsr}
\end{equation}
where $\alpha_j$ and $\beta_j=\pm 1$ and $S(\alpha,\beta)$ is an
$L\times L$ matrix given as
\begin{equation}
S_R(\alpha,\beta)=\left( \begin{array}{cccccc}
z_0 & z_{-1}&0&0&\cdot&0\\
z_1 & 0  &z_{-2}&0&\cdot&0\\ 
0   &z_2 &0&z_{-3}&\cdot&0\\
\cdot&\cdot&\cdot&\cdot&\cdot&\cdot\\
0&0&0&\cdot&0&z_{1-L}\\
0&0&0&\cdot&z_{L-1}&z_L
\end{array}\right)
\label{sr}
\end{equation}
where
\begin{equation}
z_m=q(\alpha, \beta, m|v)
\end{equation}
with
\begin{eqnarray}
q(+,\beta,m|v)&=&H(v+K+2m\eta)\tau_{\beta,m},\nonumber\\
q(-,\beta,m|v)&=&\Theta(v+K+2m\eta)\tau_{\beta,m}
\label{qtau}
\end{eqnarray} 
and the integer $L$ is defined by (\ref{root})
The $\tau_{\beta,m}$ are generically arbitrary but we note that if
they are all set equal to unity then $Q_R(v)$ is so singular that
its rank becomes 1. 

Furthermore ref. \cite{bax1} also defines
the companion matrix  $Q_L(v)$
\begin{equation}
[Q_L(v)]_{\alpha |\beta}={\rm Tr}S_L(\alpha_1, \beta_1)
S_L(\alpha_2, \beta_2)\cdots S_L(\alpha_N, \beta_N)
\label{qlsl}
\end{equation}
where $\alpha_j$ and $\beta_j=\pm 1$ and $S_L(\alpha,\beta)$ is an
$L\times L$ matrix given as
\begin{equation}
S_L(\alpha,\beta)=\left( \begin{array}{cccccc}
z'_0 & z'_{-1}&0&0&\cdot&0\\
z'_1 & 0  &z'_{-2}&0&\cdot&0\\ 
0   &z'_2 &0&z'_{-3}&\cdot&0\\
\cdot&\cdot&\cdot&\cdot&\cdot&\cdot\\
0&0&0&\cdot&0&z'_{1-L}\\
0&0&0&\cdot&z'_{L-1}&z'_L
\end{array}\right)
\label{sl}
\end{equation}
with 
\begin{equation}
z'_m=q'(\alpha, \beta, m|v)
\end{equation}
and 
\begin{eqnarray}
q(\alpha,+,m|v)&=&\tau'_{\alpha,m}H(v-K-2m\eta),\nonumber\\
q(\alpha,-,m|v)&=&\tau'_{\alpha,m}\Theta(v-K-2m\eta)
\label{qtaup}
\end{eqnarray} 

In ref. \cite{bax1} it is shown that
\begin{equation}
T(v)Q_R(v)=h^N(v-\eta)Q_R(v+2\eta)+h^N(v+\eta)Q_R(v-2\eta)
\label{tqr}
\end{equation}
and
\begin{equation}
Q_L(v)T(v)=h^N(v-\eta)Q_L(v+2\eta)+h^N(v+\eta)Q_L(v-2\eta)
\label{tql}
\end{equation}
where
\begin{equation}
h(v)=\Theta(0)\Theta(v)H(v)
\label{hvdfn}
\end{equation}
and it is further shown that
\begin{equation}
Q_L(v)Q_R(v')=Q_L(v')Q_R(v).
\end{equation} 

Thus it follows that if we define
\begin{equation}
Q_{72}(v)=Q_R(v)Q^{-1}_R(v_0)=Q^{-1}_L(v_0)Q_L(v)
\label{q72def}
\end{equation}
then
\begin{equation}
T(v)Q_{72}(v)=h^N(v-\eta)Q_{72}(v+2\eta)+h^N(v+\eta)Q_{72}(v-2\eta)
\label{tq}
\end{equation}
with
\begin{equation}
[T(v),T(v')]=[T(v),Q_{72}(v')]=[Q_{72}(v),Q_{72}(v')]=0
\label{comm}
\end{equation}

We note the periodicity relations 
\begin{eqnarray}
T(v\pm 2K)&=&(-1)^NT(v)\label{pt}\\
Q_{R,L}(v\pm 2K)&=&SQ_{R,L}(v)=Q_{R,L}(v)S\label{pqrl}\\
h(v+2K)&=&-h(v)
\label{ph}
\end{eqnarray}
where
\begin{equation}
S=\prod_{j=1}^N\sigma^z_j
\end{equation}
which are consistent with (\ref{tqr}) and (\ref{tql})
and we also note the commutation relation
\begin{equation}
Q_R(v_1)Q^{-1}_R(v_2)Q_R(v_3)=Q_R(v_3)Q^{-1}_R(v_2)Q_R(v_1)
\label{qr3rel}
\end{equation}

\section{Fusion relations and the conjectured functional equation for $Q$}

The functional equation (\ref{tq}) is derived for the matrices $T(v)$
and $Q_{72}(v)$ but it follows from the commutation relations (\ref{comm})
that the four matrices $T(v),~Q_{72}(v),~Q_{72}(v\pm 2\eta)$ may be
simultaneously diagonalised and thus (\ref{tq}) may be regarded as an
equation for the eigenvalues. The matrix $Q_{72}(v)$ is found (empirically,
we know of no mathematical proof in the literature) to be nondegenerate
and thus if $T(v)$ were also nondegenerate there would be a $1-1$ map
between eigenvalues of $T(v)$ and $Q_{72}(v).$ However, in ref. \cite{bax1}
the condition (\ref{root}) holds and in this case, when the number of
sites in the lattice $N$ is sufficiently large, the matrix $T(v)$
always has degenerate eigenvalues. There is thus a many to one map of
eigenvalues of $Q_{72}(v)$ to eigenvalues of $T(v)$ and this leads to 
the fact
that the functional equation is not sufficient to determine all the
eigenvalues of $Q_{72}(v)$ which correspond to a degenerate eigenvalue
of $T(v).$

In order to resolve this problem of degeneracy and multiplicity of
the eigenvalue of $T(v)$ we recently conjectured \cite{fm1} the following
functional equation for $Q_{72}$

 \vspace{.2in}
{\bf CONJECTURE}
\vspace{.1in}

For $N$ even and either $L$ even or $L$ and $m_1$ odd  
\begin{eqnarray}
& &e^{-N\pi i v/2K}Q_{72}(v-iK')\nonumber\\
&=&A\sum_{l=0}^{L-1}h^N(v-(2l+1)m_1K/L)
{Q_{72}(v)\over Q_{72}(v-2lm_1K/L)Q_{72}(v-2(l+1)m_1K/L)}
\label{con}
\end{eqnarray}
where $A$ is a normalizing  constant matrix independent of $v$ 
that commutes with $Q_{72}(v).$ What this matrix is depends
on the normalization value of $v_0$ in the definition (\ref{q72def}) 
of $Q_{72}(v)$.  

\vspace {.1in}

We prove this conjecture for $L=2$ in sec. 7 and have numerically
verified it for $L=3$ for various values of $N.$

The matrix $Q_{72}(v)$ as defined by (\ref{q72def}) is not in the form of
$Q_R(v)$ and $Q_L(v)$ of being the trace of a product of matrices and
thus it is natural to rewrite the conjecture (\ref{con}) in 
terms of $Q_L(v)$ and
$Q_R(v)$ as
\begin{eqnarray}
&&e^{-N\pi iv/2K}Q_L(v-iK')=Q_L(v_0)AQ_R(v_0)\nonumber\\
&&\times\sum_{l=0}^{L-1}h^N(v-(2l+1)m_1K/L)
Q_R^{-1}(v-2lm_1K/L)Q_R(v)Q_R^{-1}(v-2(l+1)m_1K/L)
\label{con2}
\end{eqnarray}
The form (\ref{con2}) of the conjecture 
is strikingly similar
in form to the functional equation 4.40 in ref. \cite{bbp} 
of the chiral Potts model 
if we make the identification of $Q_L(v)$ with ${\hat T}_{cp}$ and
$Q_R(v)$ with $T_{cp}$ where the subscript cp indicates the quantities
in ref. \cite{bbp}. It is therefore natural to search
for a proof of (\ref{con}) by following the methods of
ref. \cite{bbp}. 

We begin by writing (\ref{tqr}) in the form
\begin{equation}
T(v)=h^N(v-\eta)Q_R(v+2\eta)Q^{-1}_R(v)
+h^N(v+\eta)Q_R(v-2\eta)Q^{-1}_R(v)
\label{help1}
\end{equation}
which is analogous to (4.20) of ref. \cite{bbp} if we identify $T(v)$
with $\tau^{(2)}_{cp}$. 

To continue we need to define quantities analogous to
$\tau^{(j)}_{cp}$ which obey functional equations analogous to (4.27a)
of ref. \cite{bbp}. The appropriate objects are the ``fusion matrices''
$T^{(j)}(v)$ which may be defined recursively, for any $\eta$ not just
(\ref{root}), by
\begin{eqnarray}
&&T^{(2)}(v)=T(v)\label{fus0}\\
&&T^{(2)}(v-2\eta)T^{(2)}(v-4\eta)=h^N(v-3\eta)T^{(3)}(v-4\eta)
+h^N(v-\eta)h^N(v-5\eta)
\label{fus1}
\end{eqnarray}
and for $j\geq 3$
\begin{equation}
T^{(2)}(v-2\eta)T^{(j)}(v-2j\eta)=h^N(v-3\eta)T^{(j+1)}(v-2j\eta)
+h^N(v-\eta)T^{(j-1)}(v-2j\eta).
\label{fus2}
\end{equation}

From these defining equations we show that $T^{(j)}$ may be written in
terms of  $Q_R(v)$ as
\begin{eqnarray}
&&T^{(j)}(v-2(j-1)m_1K/L)=\sum_{l=0}^{j-1}h^N(v-(2l+1)m_1K/L)\nonumber\\
&&\times Q_R(v)Q^{-1}_R(v-2lm_1K/L)
Q_R(v-2jm_1K/L)Q_R^{-1}(v-2(l+1)m_1K/L)\nonumber\\
\label{analogue2}
\end{eqnarray}
by directly substituting (\ref{analogue2}) into (\ref{fus1}) and
(\ref{fus2}), using the commutation relation (\ref{qr3rel})
and noting that (\ref{analogue2}) reduces to (\ref{help1}) if we 
set $j=2$ and send $v\rightarrow v+2\eta.$
Equation (\ref{analogue2}) is the analogue of 4.34 of ref. \cite{bbp}

Now define
\begin{equation}
M=Q_L(v_0)AQ_R(v_0)
\label{mqdef}
\end{equation}
and multiply (\ref{con2}) on the left by $Q_R(v)M^{-1}$ we obtain
\begin{eqnarray}
& &e^{-N\pi i v/2K}Q_R(v)M^{-1}Q_L(v-iK')\nonumber \\
&=&\sum_{l=0}^{L-1}h^N(v-(2l+1)m_1K/L)Q_R(v)
Q_R^{-1}(v-2lm_1K/L)Q_R(v)Q^{-1}_R(v-2(l+1)\eta)
\label{con3}
\end{eqnarray}
If we now use the periodicity
property (\ref{pqrl}) in the right hand side of (\ref{analogue2})
with $j=L$ and compare with the right hand side of the conjectured
functional equation (\ref{con3}) we conclude that the conjectured
functional equation will hold if we can prove that
\begin{equation}
T^{(L)}(v-2(L-1)m_1K/L)=e^{-iN\pi v/2K}Q_R(v)M^{-1}Q_L(v-iK')S^{m_1}.
\label{closing}
\end{equation}
which is the analogue of 4.44 of ref. \cite{bbp}

We conclude this section by noting that
in the chiral Potts model there is a functional equation (4.27c) of
ref. \cite{bbp} which relates $\tau^{(L+1)}$ to $\tau^{(L-1)}$. 
To obtain the analogous equation in the eight vertex model we specialize
$j=L+1$  in (\ref{analogue2}) and use (\ref{pqrl}) to get
\begin{eqnarray}
&&T^{(L+1)}(v-2m_1K)=\sum_{l=0}^{L}h^N(v-(2l+1)m_1K/L)\nonumber\\
&&Q_R(v)Q^{-1}_R(v-2lm_1K/L)
Q_R(v-2m_1K/L)S^{m_1}Q_R^{-1}(v-2(l+1)m_1K/L)\nonumber\\
\label{fun1}
\end{eqnarray}
We now write the terms $l=0,L$ separately and use (\ref{ph}) to find
\begin{eqnarray}
&&T^{(L+1)}(v-2m_1K)\nonumber\\
&&=[1+(-1)^N]h^N(v-m_1K/L)S^{m_1}
+\sum_{l=1}^{L-1}h^N(v-(2l+1)m_1K/L)\nonumber\\
&&\times Q_R(v)Q^{-1}_R(v-2lm_1K/L)
Q_R(v-2m_1K/L)S^{m_1}Q_R^{-1}(v-2(l+1)m_1K/L).\nonumber\\
\label{fun2}
\end{eqnarray}
In the sum we use the commutation relation (\ref{qr3rel}), set
$l=k+1$ and use (\ref{pqrl}) to write 
$S^{m_1}Q_R(v)=Q_R(v-2m_1K)=Q_R(v-2m_1K/L-2(L-1)m_1K/L)$ 
and thus we find
\begin{eqnarray}
&&T^{(L+1)}(v-2m_1K)\nonumber\\
&&=[1+(-1)^N]h^N(v-m_1K/L)S^{m_1}
+\sum_{k=0}^{L-2}h^N(v-2m_1K/L-(2k+1)m_1K/L)\nonumber\\
&&\times Q_R(v-2m_1K/L)
Q^{-1}_R(v-2m_1K/L-2km_1K/L)
Q_R(v-2m_1K/L-2(L-1)m_1K/L)\nonumber\\
&&\times Q_R^{-1}(v-2m_1K/L-2(k+1)m_1K/L).\nonumber\\
\label{fun3}
\end{eqnarray}
The sum on the right hand side is seen to be
$T^{(L-1)}(v-2m_1K/L-2(L-2)m_1K/L)$ by use of (\ref{analogue2}) and thus
\begin{equation}
T^{(L+1)}(v-2m_1K)=[1+(-1)^N]h^N(v-m_1K/L)S^{m_1}
+T^{(L-1)}(v+2m_1K/L-2m_1K)
\label{fun4}
\end{equation}
 and finally by use of (\ref{pt}) we obtain the desired result
\begin{equation}
T^{(L+1)}(v)=[1+(-1)^N]h^N(v-m_1K/L)S^{m_1}
+T^{(L-1)}(v+2m_1K/L)
\label{fun5}
\end{equation}

In the six vertex limit this functional equation was first exhibited
by Nepomechie as (1.3) of ref. \cite{nep1} where it is proven for $L=2,3,4$ 
and used to study various open six vertex chains at roots of unity in
\cite{nep1}-\cite{nep3}. For the eight vertex model this result seems to have
been first obtained by Baxter \cite{bax3}.

\section{Explicit fusion matrices for $j=3,4,5$}

We introduced the matrices $T^{(j)}(v)$ by the recursion relations 
(\ref{fus0})-(\ref{fus2})
and called them ``fusion matrices''. However in ref. \cite{bbp} the
analogous matrices $\tau^{(j)}$ are not defined by the analogue of 
(\ref{fus0})-(\ref{fus2})
but rather they are defined as the trace of products of explicitly
given $j\times j$ matrices and (\ref{fus0})-(\ref{fus2}) follows as a theorem. 
The explicit form of $\tau^{(j)}$ is used in the proof of the analogue
of (\ref{closing}). 

The existence of matrices  $T^{(j)}(v)$ written in the form
\begin{equation}
T^{(j)}(v)|_{\mu,\nu}={\rm Tr}R^{(j)}(\mu_1,\nu_2)(v)\cdots 
R^{(j)}(\mu_N,\nu_N)(v)
\label{trj}
\end{equation}
with $R^{(j)}(\mu.\nu)$ 
being $j \times j$ matrices which generalizes (\ref{t8}) and has
the global property that $[T^{(j)}(v),T^{(j)}(v')]=0$ 
by virtue of $R^{(j)}(\mu,\nu)(v)$ and $R^{(j)}(\mu,\nu)(v')$ 
satisfying a local
Yang-Baxter equation using
\begin{equation}
R^{(2)}(\mu,\nu)(v)=W_8(\mu,\nu)
\label{bwr2}
\end{equation}
as the elements of the $2\times 2$ intertwining  matrix has been 
extensively studied
\cite{kure}-\cite{djkmo} for both the 8 and the 6 vertex model.
The final result of these studies for the 8 vertex model is given by
lemma 2.3.1 of ref. \cite{djkmo} which says  that $T^{(j)}(v)$ as defined by 
(\ref{trj}) is given in terms of a matrix
$R^{(j)}(\mu,\nu)(v)$  which is constructed from $R^{(2)}(\mu,\nu)(v)$ by 
\begin{equation}
R^{(j)}(\mu,\nu)(v)={P\sum_{\nu_1,\cdots
\nu_{j-2}}R^{(2)}(\mu,\nu_1)(v)
R^{(2)}(\nu_1,\nu_2)(v+2\eta)\cdots
R^{(2)}(\nu_{j-2},\mu)(v+2(j-2)\eta)
\over \prod_{l=0}^{j-3}h[v+(2l+1)\eta]}
\label{fusionrule}
\end{equation}
where $P$ is the projection from the internal space of dimension
$2^{j-1}$ to the space of dimension $j$ of completely symmetric
tensors. This construction is known as ``fusion''. The matrix 
$R^{(j)}(v)$ has no poles and $T^{(j)}(v)$ satisfies (\ref{fus0})-(\ref{fus2}).

Unfortunately the result (\ref{fusionrule}) is not as explicit as the
expression for $\tau^{(j)}$ in the chiral Potts model \cite{bbp} or
the corresponding expression for the fusion weights in the RSOS model
\cite{djkmo}. However, it would appear that such an explicit form
would be of help in proving the conjecture (\ref{closing}) for the 8
vertex model. Therefore in order to gain insight into the fusion
matrices we have constructed the matrices $T^{(j)}(v)$ directly from
(\ref{fusionrule})  for $j=3,4,5$. The computation is straightforward
and makes extensive use of properties of theta functions (presented in
detail for example in ref.\cite{w}). In particular we use
 the two addition formulas for theta
functions (15.4.25) and (15.4.26) of ref. \cite{baxb}
\begin{eqnarray}
&&\Theta(u)\Theta(v)\Theta(a-u)\Theta(a-v)-H(u)H(v)H(a-u)H(a-v)\nonumber\\
&&=\Theta(0)\Theta(a)\Theta(u-v)\Theta(a-u-v)
\label{ad1}
\end{eqnarray}
and
\begin{eqnarray}
&&H(v)H(a-v)\Theta(u)\Theta(a-u)-\Theta(v)\Theta(a-v)H(u)H(a-u)\nonumber\\
&&=\Theta(0)\Theta(a)H(v-u)H(a-u-v)
\label{ad2}
\end{eqnarray}
and the fact that in the set of functions given for $a$ fixed and $j$
an integer 
\begin{equation}
\Theta(v+j\eta)H(v+(a-j)\eta)
\end{equation}  
only two are linearly independent.
Similarly of the functions for fixed $a$
\begin{equation}
\Theta(v+j\eta)\Theta(v+(a-j)\eta),~~H(v+j\eta)H(v+(a-j)\eta)
\end{equation}
only two are linearly independent. There are accordingly many
equivalent ways to write the theta functions in
$R^{(j)}(\mu,\nu)(v)$. 

We note that the matrix $R^{(j)}(-,-)(v)$ 
is obtained from $R^{(j)}(+,+)(v)$ and
the matrix $R^{(j)}(-,+)(v)$ is obtained from $R^{(j)}(+,-)(v)$ by the
interchange $\Theta(v+2k\eta)\leftrightarrow H(v+2k\eta).$
Furthermore the matrix elements have the symmetry property
\begin{eqnarray}
&&R^{(j)}(\mu,\nu)(v)_{m,n}\leftrightarrow
R^{(j)}(\mu,\nu)(v)_{j-1-m,j-1-n}\nonumber\\
&&{\rm by~the ~substitution}~\Theta(v+2a\eta)\leftrightarrow H(v+2a\eta)
\label{symmetry}
\end{eqnarray}

With these provisos we have the following results:

\subsection{The matrices $R^{(3)}(\mu,\nu)(v)$}
The matrix $R^{(3)}(+,+)(v)$ is
\begin{eqnarray}
& &R^{(3)}(+,+)(v)=\nonumber\\
& &\left( \begin{array}{ccc}
{\Theta^2(2\eta)\over \Theta(0)}\Theta(v-\eta)H(v+3\eta)&0
&{H^2(2\eta)\over \Theta(0)}H(v-\eta)\Theta(v+3\eta)\\
0&\Theta(4\eta)H(v+\eta)\Theta(v+\eta)&0\\
{H^2(2\eta)\over \Theta(0)}\Theta(v-\eta)H(v+3\eta)&0
&{\Theta^2(2\eta)\over \Theta(0)}H(v-\eta)\Theta(v+3\eta)
\label{w31}
\end{array}\right)
\end{eqnarray}
and the  matrix $R^{(3)}(+,-)(v)$ is
\begin{eqnarray}
& &R^{(3)}(+,-)(v)=\nonumber\\
& &\left( \begin{array}{ccc}
0&H(4\eta)H^2(v+\eta)&0\\
{\Theta(2\eta)H(2\eta)\over \Theta(0)}\Theta(v-\eta)\Theta(v+3\eta)&
0&{H(2\eta)\Theta(2\eta)\over \Theta(0)}H(v-\eta)H(v+3\eta)\\
0&H(4\eta)\Theta^2(v+\eta)&0
\label{w33}
\end{array}\right).
\end{eqnarray}

\subsection{The matrices $R^{(4)}(\mu,\nu)(v)$}

The matrix $R^{(4)}(+,+)(v)$ is 
\begin{equation}
R^{(4)}(+,+)(v)=
\left(\begin{array}{cccc}
R^{(4)}_{00}&0&R^{(4)}_{02}&0\\
0&R^{(4)}_{11}&0&R^{(4)}_{13}\\
R^{(4)}_{20}&0&R^{(4)}_{22}&0\\
0&R^{(4)}_{31}&0&R^{(4)}_{33}\\
\end{array}\right)
\end{equation}
where the 4 independent 
elements of $R^{(4)}(++)$ are
\begin{eqnarray}
R^{(4)}_{00}(+,+)&=&{\Theta^3(2\eta)\over \Theta^2(0)}\Theta(v-\eta)H(v+5\eta)\label{400++}\\
R^{(4)}_{20}(+,+)&=&{H^2(2\eta)\Theta(2\eta)\over \Theta^2(0)}
\Theta(v-\eta)H(v+5\eta)
\label{420++}\\
R^{(4)}_{11}(+,+)&=&{\Theta(4\eta)H(4\eta)\over H(2\eta)}
H(v+\eta)\Theta(v+3\eta)-{\Theta^3(2\eta)\over \Theta^2(0)}
H(v-\eta)\Theta(v+5\eta)
\label{411++}\\
R_{31}^{(4)}(+,+)&=&{H^2(4\eta)\over
\Theta(2\eta)}\Theta(v+\eta)H(v+3\eta)
-{H^2(2\eta)\Theta(2\eta)\over \Theta^2(0)}H(v-\eta)\Theta(v+5\eta)
\label{431++}
\end{eqnarray}
and the remaining 4 elements are obtained by the symmetry
(\ref{symmetry}).

The matrix $R^{(4)}(+,-)(v)$ is
\begin{equation}
R^{(4)}(+,-)(v)=
\left(\begin{array}{cccc}
0&R^{(4)}_{01}&0&R^{(4)}_{03}\\
R^{(4)}_{10}&0&R^{(4)}_{12}&0\\
0&R^{(4)}_{21}&0&R^{(4)}_{23}\\
R^{(4)}_{30}&0&R^{(4)}_{32}&0\\
\end{array}\right)
\label{r4form+-}
\end{equation}
where the 4 independent elements are
\begin{eqnarray}
R^{(4)}_{10}(+,-)&=&{\Theta^2(2\eta)H(2\eta)\over \Theta^2(0)}
\Theta(v-\eta)\Theta(v+5\eta)
\label{410+-}\\
R^{(4)}_{30}(+,-)&=&{H^3(2\eta)\over \Theta^2(0)}
\Theta(v-\eta)\Theta(v+5\eta)\label{430+-}\\
R^{(4)}_{21}(+,-)&=&-{H^3(2\eta)\Theta(4\eta)\over
\Theta^3(0)}H(v+\eta)H(v+3\eta)
+{\Theta^3(2\eta)H(4\eta)\over \Theta^3(0)}\Theta(v+\eta)\Theta(v+3\eta)
\label{421+-}\\
&=&{\Theta^2(4\eta)H(4\eta)\over \Theta(0) \Theta(6\eta)}\Theta(v-\eta)
\Theta(v+5\eta)+{\Theta(2\eta)H^2(2\eta)H(6\eta)\over \Theta^2(0)\Theta(6\eta)}
H(v-\eta)H(v+5\eta)\label{n421+-}\\
R^{(4)}_{01}(+,-)&=&-{\Theta^2(2\eta)H(2\eta)\over \Theta^2(0)}
H(v-\eta)H(v+5\eta)
+{H^2(4\eta)\over H(2\eta)}H(v+\eta)H(v+3\eta)
\label{401+-}\nonumber\\
&=&{H^3(4\eta)\over 
\Theta(0)\Theta(6\eta)}\Theta(v-\eta)\Theta(v+5\eta)
+{\Theta^3(2\eta)H(6\eta)\over \Theta^2(0) \Theta(6\eta)}H(v-\eta)H(v+5\eta)
\label{n401+-}
\end{eqnarray}
and the four other elements obtained by the symmetry (\ref{symmetry})

\subsection{The matrices $R^{(5)}(\mu.\nu)(v)$}

The matrix $R^{(5)}(+,+)(v)$ is
\begin{equation}
R^{(5)}(+,+)(v)=
\left(\begin{array}{ccccc}
R^{(5)}_{00}&0&R^{(5)}_{02}&0&R^{(5)}_{04}\\
0&R^{(5)}_{11}&0&R^{(5)}_{13}&0\\
R^{(5)}_{20}&0&R^{(5)}_{22}&0&R^{(5)}_{24}\\
0&R^{(5)}_{31}&0&R^{(5)}_{33}&0\\
R^{(5)}_{40}&0&R^{(5)}_{42}&0&R^{(5)}_{44}\\
\end{array}\right)
\end{equation}
where the 6 independent non symmetric elements are
\begin{eqnarray}
&&R^{(5)}_{00}(+,+)
={\Theta^4(2\eta)\over \Theta^3(0)}\Theta(v-\eta)H(v+7\eta)\\
&&R^{(5)}_{20}(+,+)= {H^{2}(2\eta)\Theta^{2}(2\eta)\over \Theta^3(0)}
\Theta(v-\eta)H(v+7\eta)\\
&&R^{(5)}_{40}(+,+)={H^4(2\eta)\over \Theta^3(0)}\Theta(v-\eta)H(v+7\eta)\\
&&R^{(5)}_{11}(+,+)=
{\Theta^{2}(2\eta)\Theta^{2}(4\eta)\over \Theta^3(0)}\Theta(v+\eta)H(v+5\eta)
-{H^{2}(2\eta)H^{2}(4\eta)\over \Theta^3(0)}H(v+\eta)\Theta(v+5\eta)\\
&&R^{(5)}_{31}(++)= 
{\Theta^{2}(2\eta)H^{2}(4\eta)\over \Theta^3(0)}\Theta(v+\eta)H(v+5\eta)-
{H^{2}(2\eta)\Theta^{2}(4\eta)\over \Theta^3(0)}H(v+\eta)\Theta(v+5\eta)\\
&&R^{(5)}_{02}(+,+)=\nonumber \\
&&-\frac{H^{2}(2\eta)
\Theta^{2}(2\eta)H(8\eta)}{\Theta^2(0)H(4\eta)\Theta(4\eta)}
H(v+3\eta)\Theta(v+3\eta)+\frac{H^{3}(4\eta)}
{H(2\eta)\Theta(2\eta)}
H(v+\eta)\Theta(v+5\eta),
\end{eqnarray}
the companion non symmetric elements are obtained by the symmetry 
(\ref{symmetry})
and the one symmetric element is
\begin{equation}
R^{(5)}_{22}(+,+)=
\Theta^{-3}(0)\left(\Theta^{3}(2\eta)\Theta(6\eta)-
H^{3}(2\eta)H(6\eta)\right)
H(v+3\eta)\Theta(v+3\eta)
\end{equation}

The matrix $R^{(5)}(+,-)(v)$ is
\begin{equation}
R^{(5)}(+,-)(v)=
\left(\begin{array}{ccccc}
0&R^{(5)}_{01}&0&R^{(5)}_{03}&0\\
R^{(5)}_{10}&0&R^{(5)}_{12}&0&R^{(5)}_{14}\\
0&R^{(5)}_{21}&0&R^{(5)}_{23}&0\\
R^{(5)}_{30}&0&R^{(5)}_{32}&0&R^{(5)}_{34}\\
0&R^{(5)}_{41}&0&R^{(5)}_{43}&0\\
\end{array}\right)
\end{equation}
where the six independent elements are
\begin{eqnarray}
&&R^{(5)}_{10}(+,-) =  
{H(2\eta)\Theta^{3}(2\eta)\over \Theta^3(0)}\Theta(v-\eta)\Theta(v+7\eta)\\
&&R^{(5)}_{30}(+,-) =  
{H^{3}(2\eta)\Theta(2\eta)\over \Theta^3(0)}\Theta(v-\eta)\Theta(v+7\eta)\\
&&R^{(5)}_{01}(+,-) =  
\frac{\Theta(2\eta)}{\Theta^2(0)\Theta(6\eta)}
[H^{3}(4\eta)\Theta(v-\eta)\Theta(v+7\eta)
+\Theta^{2}(2\eta)H(8\eta)H(v+\eta)H(v+5\eta)]\\
&&R^{(5)}_{21}(+,-) = {H(4\eta)\Theta(4\eta)\over \Theta^3(0)} 
[\Theta^{2}(2\eta)\Theta(v+\eta)\Theta(v+5\eta)-
H^{2}(2\eta)H(v+\eta)H(v+5\eta)]\\
&&R^{(5)}_{41}(+,-) =  
\frac{H(2\eta)}{\Theta^2(0)H(6\eta)}
[H^{3}(4\eta)\Theta(v-\eta)\Theta(v+7\eta)
+H^{2}(2\eta)H(8\eta)\Theta(v+\eta)\Theta(v+5\eta)]\\
&&R^{(5)}_{12}(+,-) = 
{H(2\eta)\Theta(6\eta)\over \Theta(0)}H(v+\eta)H(v+5\eta)
+ \frac{H(2\eta)\Theta^{3}(2\eta)H(8\eta)}
{\Theta^2(0)H(4\eta)\Theta(4\eta)}H^{2}(v+3\eta)
\end{eqnarray}
and the other six elements are obtained by the  symmetry (\ref{symmetry}).

\section{A similarity transformation}

The fusion matrices derived in the preceding section by direct
application of the fusion construction do not 
have a particularly revealing form. Furthermore they do not
directly  reveal the reduction (\ref{fun5}) of $T^{(L+1)}(v)$ 
at the root of unity point (\ref{root}).
 However, because the form (\ref{trj}) for $T^{(j)}(v)$  is the trace of a
product of matrices any similarity transformation of the
$R^{(j)}(\mu,\nu)(v)$ 
\begin{equation}
{\tilde R}^{(j)}(\mu,\nu)(v)=M_j(\eta)R^{(j)}(\mu,\nu)(v)M_j^{-1}(\eta)
\end{equation}
with $M_j(\eta)$ independent of $v$ is just as good for our purposes
as the original $R^{(j)}(\mu,\nu)(v).$ 

The form of the functional equation (\ref{fun5}) 
will follow if we can determine a
similarity transformation matrix $M_j(\eta)$ such that when the root of
unity condition (\ref{root}) holds the matrix
elements of ${\tilde R}^{(L+1)}(\mu,\nu)$ have the property that
\begin{equation}
{\tilde R}^{(L+1)}(\mu,\nu)_{0,k}={\tilde
R}^{(L+1)}(\mu,\nu)_{L-1,k}=0~~{\rm for}~1\leq k \leq L-2
\label{decom}
\end{equation}
The matrix $R^{(3)}(\mu,\nu)$ already has this property but the matrices
$R^{(4)}(\mu,\nu)$ and $R^{(5)}(\mu,\nu)$ do not. However, because the
functional equation (\ref{fun5}) has been proven true in sec. 3 by an
independent method it must be possible to find a similarity
transformation which does in fact put $R^{(4)}(\mu,\nu)$ and
$R^{(5)}(\mu,\nu)$ into the required  form. It is straightforward to
determine these matrices. In fact these matrices are not unique and
have several arbitrary parameters we can freely chose. We have
determined the families of these similarity transformations for $L=3$
and $4.$ These similarity
transformations may then be extended  from the root of unity case to  
the case of arbitrary $\eta$ essentially by replacing $m_1K/L$
everywhere by $\eta.$ When this is done we find the following
similarity transformations which are easily verified.

\subsection{Transformation of $R^{(4)}(\mu,\nu)$}

The transformation matrix is
\begin{equation}
M_4(\eta)=\left(\begin{array}{cccc}
1&0&-f&0\\
0&1&0&0\\
0&0&1&0\\
0&-f&0&1\\
\end{array}\right)
\end{equation}
and 
\begin{equation}
M_4^{-1}(\eta)=\left(\begin{array}{cccc}
1&0&f&0\\
0&1&0&0\\
0&0&1&0\\
0&f&0&1\\
\end{array}\right)
\end{equation}
with 
\begin{equation}
f=H^2(2\eta)/\Theta^2(2\eta)
\end{equation}

Thus using the notation
\begin{eqnarray}
\Theta_a&=&\Theta(v+a\eta)\\
H_a&=&H(v+a\eta)
\end{eqnarray}
we have
\begin{equation}
{\tilde R}^{(4)}(+,+)=M_4(\eta)R^{(4)}(++)M_4^{-1}(\eta)
=\left(\begin{array}{cccc}
C^{(4)}_{00}\Theta_{-1}H_5&0&C^{(4)}_{02}\Theta_3H_1&0\\
0&C^{(4)}_{11}\Theta_1H_3&0&C^{(4)}_{13}\Theta_5H_{-1}\\
C^{(4)}_{20}\Theta_{-1}H_5&0&C^{(4)}_{22}\Theta_3H_1&0\\
0&C^{(4)}_{31}\Theta_1H_3&0&C^{(4)}_{33}\Theta_5H_{-1}\\
\end{array}\right)
\label{tr4++}
\end{equation}
where
\begin{eqnarray}
C^{(4)}_{00}&=&C^{(4)}_{33}={\Theta(0)\Theta(4\eta)\over\Theta(2\eta)}\\
C^{(4)}_{22}&=&C^{(4)}_{11}={\Theta^2(4\eta)\over\Theta(2\eta)}\\
C^{(4)}_{20}&=&C^{(4)}_{13}={H^2(2\eta)\Theta(2\eta)\over \Theta^2(0)}\\
C^{(4)}_{31}&=&C^{(4)}_{02}={\Theta^2(0)H(2\eta)H(6\eta)\over \Theta^3(2\eta)}
\end{eqnarray}

and
\begin{equation}
{\tilde R}^{(4)}(+,-)=M_4(\eta)R^{(4)}(+,-)M_4^{-1}(\eta)
=\left(\begin{array}{cccc}
0&C^{(4)}_{01}H_1H_3&0&0\\
C^{(4)}_{10}\Theta_{-1}\Theta_7&0&C^{(4)}_{12}H_1H_3&0\\
0&C^{(4)}_{21}\Theta_1\Theta_3&0&C^{(4)}_{23}H_{-1}H_5\\
0&0&C^{(4)}_{32}\Theta_1\Theta_3&0\\
\end{array}\right)
\label{tr4+-}
\end{equation}
with
\begin{eqnarray}
C^{(4)}_{10}&=&C^{(4)}_{23}={\Theta^2(2\eta)H(2\eta)\over \Theta^2(0)}\\
C^{(4)}_{21}&=&C^{(4)}_{12}={H(4\eta)\Theta(4\eta)\over \Theta(2\eta)}\\
C^{(4)}_{32}&=&C^{(4)}_{01}={\Theta^2(0)H(6\eta)\over \Theta^2(2\eta)}
\end{eqnarray}

\subsection{Transformation of $R^{(5)}(\mu,\nu)$}

The transformation matrix is
\begin{equation}
M_5(\eta)=\left(\begin{array}{ccccc}
1&0&f&0&0\\
0&1&0&g&0\\
0&0&h&0&0\\
0&g&0&1&0\\
0&0&f&0&1\\
\label{5sdef}
\end{array}\right)
\end{equation}
 where
\begin{eqnarray}
f&=&-{H^2(4\eta)\over \Theta^2(4\eta)}\label{sfdef}\\
g&=&-{H^2(2\eta)\over \Theta^2(2\eta)}\label{sgdef}\\
h&=&{\Theta^2(0)H(2\eta)\over H(4\eta)\Theta(4\eta)\Theta(2\eta)}\label{shdef}
\end{eqnarray}
and 
\begin{equation}
M_5^{-1}(\eta)=\left(\begin{array}{ccccc}
1&0&a&0&0\\
0&b&0&c&0\\
0&0&d&0&0\\
0&c&0&b&0\\
0&0&a&0&1\\
\label{5sidef}
\end{array}\right)
\end{equation}
with
\begin{eqnarray}
a&=&-{f\over h}={H^3(4\eta)\Theta(2\eta)\over
\Theta(4\eta)\Theta^2(0)H(2\eta)}\\
b&=&{1\over 1-g^2}={\Theta^4(2\eta)\over \Theta^3(0)\Theta(4\eta)}\\
c&=&-{g\over 1-g^2}=
{H^2(2\eta)\Theta^2(2\eta)\over\Theta^3(0)\Theta(4\eta)}\\
d&=&{1\over h}={H(4\eta)\Theta(4\eta)\Theta(2\eta)\over H(2\eta)\Theta^2(0)}
\end{eqnarray}

Thus we find
\begin{equation}
{\tilde R}^{(5)}(+,+)=M_5(\eta)R^{(5)}(++)M_5(\eta)^{-1}
\left(\begin{array}{ccccc}
C^{(5)}_{00}\Theta_{-1}H_7&0&C^{(5)}_{02}\Theta_3H_3
&0&C^{(5)}_{04}\Theta_7H_{-1}\\
0&C^{(5)}_{11}\Theta_1H_5&0&C^{(5)}_{13}\Theta_5H_1&0\\
C^{(5)}_{20}\Theta_{-1}H_7&0&C^{(5)}_{22}\Theta_3H_3&0
&C^{(5)}_{24}\Theta_7H_{-1}\\
0&C^{(5)}_{31}\Theta_1H_5&0&C^{(5)}_{33}\Theta_5H_1&0\\
C^{(5)}_{40}\Theta_{-1}H_7&0&C^{(5)}_{42}\Theta_3H_3&0
&C^{(5)}_{44}\Theta_7H_{-1}\\
\end{array}\right)
\label{tr5++}
\end{equation}
where
\begin{eqnarray}
C^{(5)}_{00}&=&C^{(5)}_{44}={\Theta^3(2\eta)\Theta(6\eta)\over 
\Theta(0)\Theta^2(4\eta)}\\
C^{(5)}_{20}&=&C^{(5)}_{24}={H^3(2\eta)\Theta(2\eta)
\over \Theta(0)\Theta(4\eta)H(4\eta)}\\
C^{(5)}_{40}&=&C^{(5)}_{04}
={H^3(2\eta)H(6\eta)\over \Theta(0)\Theta^2(4\eta)}\\
C^{(5)}_{02}&=&C^{(5)}_{42}={H(8\eta)H(6\eta)
\Theta(6\eta)\Theta^2(2\eta)\over \Theta^4(4\eta)}\\
C^{(5)}_{11}&=&C^{(5)}_{33}={\Theta(2\eta)\Theta(6\eta) \over \Theta(0)}\\
C^{(5)}_{13}&=&C^{(5)}_{31}={H(2\eta)H(6\eta)\over \Theta(0)}\\
C^{(5)}_{22}&=&{1\over \Theta(0) \Theta^2(4\eta)}
\left(\Theta^2(2\eta)\Theta^2(6\eta)+H^2(2\eta)H^2(6\eta)\right)
\end{eqnarray}

and  
\begin{equation}
{\tilde R}^{(5)}(+,-)=M_5(\eta)R^{(5)}(+-)M_5(\eta)^{-1}
=\left(\begin{array}{ccccc}
0&C^{(5)}_{01}H_1H_5&0&0&0\\
C^{(5)}_{10}\Theta_{-1}\Theta_7&0&C^{(5)}_{12}H_3^2&0&0\\
0&C^{(5)}_{21}\Theta_1\Theta_5&0&C^{(5)}_{23}H_1H_5&0\\
0&0&C^{(5)}_{32}\Theta^2_3&0&C^{(5)}_{34}H_{-1}H_7\\
0&0&0&C^{(5)}_{43}\Theta_1\Theta_5&0\\
\end{array}\right)
\label{tr5+-}
\end{equation}
where
\begin{eqnarray}
C^{(5)}_{10}&=&C^{(5)}_{34}={H(2\eta)\Theta(4\eta)\over
\Theta(2\eta)}\\
C^{(5)}_{21}&=&C^{(5)}_{23}={H(2\eta)\Theta(2\eta) \over \Theta(0)}\\
C^{(5)}_{32}&=&C_{12}={H(6\eta)H(4\eta)\Theta(6\eta)\over
H(2\eta)\Theta(2\eta)}\\
C^{(5)}_{43}&=&C^{(5)}_{01}
={H(8\eta)\Theta^2(2\eta)\over\Theta^2(4\eta)}\\
\end{eqnarray}

\subsection{The matrix $R^{(3)}(\mu,\nu)$}

For comparison we write $R^{(3)}(\mu,\nu)$ in the notation used above as
\begin{equation}
R^{(3)}(v)(+,+)=
\left( \begin{array}{ccc}
C^{(3)}_{00}\Theta_0H_2&0&C^{(3)}_{02}\Theta_2H_0\\
0&C^{(3)}_{11}\Theta_1H_1&0\\
C^{(3)}_{20}\Theta_0H_2&0&C_{22}^{(3)}\Theta_2H_0
\end{array}\right),
\label{tr3++}
\end{equation}
where
\begin{eqnarray}
C^{(3)}_{00}&=&C^{(3)}_{22}=\Theta^2(2\eta)/\Theta(0)\\
C^{(3)}_{20}&=&C^{(3)}_{02}=H^2(2\eta)/\Theta(0)\\
C^{(3)}_{11}&=&\Theta(4\eta)
\end{eqnarray}

\begin{equation}
R^{(3)}(v)(+,-)=
\left( \begin{array}{ccc}
0&C^{(3)}_{01}H^2_1&0\\
C^{(3)}_{10}\Theta_0\Theta_2&
0&C^{(3)}_{12}H_0H_2\\
0&C^{(3)}_{21}\Theta^2_1&0
\end{array}\right),
\label{tr3+-}
\end{equation}
where
\begin{eqnarray}
C^{(3)}_{10}&=&C^{(3)}_{12}=H(2\eta)\Theta(2\eta)/\Theta(0)\\
C^{(3)}_{01}&=&C^{(3)}_{21}=H(4\eta)
\end{eqnarray}

\subsection{Comments}

In the matrix $R^{(3)}(\mu,\nu)$ of (\ref{tr3+-}), the coefficients
$C^{(3)}_{01}=C^{(3)}_{21}$ vanish for $\eta=K/2;$
in the matrix $R^{(4)}(\mu,\nu)$ of (\ref{tr4++}),(\ref{tr4+-}), 
the coefficients
$C^{(4)}_{02}=C^{(4)}_{31}$ and  $C^{(4)}_{01}=C^{(4)}_{32}$ 
vanish for $\eta=m_1K/3;$
and in the matrix $R^{(5)}(\mu,\nu)$ of (\ref{tr5++}),(\ref{tr5+-}), 
the coefficients
$C^{(5)}_{01}=C^{(5)}_{43}$ and $C^{(5)}_{02}=C^{(5)}_{42}$
 vanish for $\eta=m_1K/4.$ Therefore the decomposition property
(\ref{decom}) holds and it is now a simple matter to see that 
the functional equation (\ref{fun5}) holds. This, of course,
was the criteria used to obtain the similarity transformation matrices
in the first place. 

What is not automatically guaranteed by our construction is that the 
transformed matrices ${\tilde R}^{(j)}(\mu,\nu)$ have matrix elements
which depend only on one of the two linearly independent products of
theta functions and that with the exception of $C^{(5)}_{22}$ all
$C^{(j)}_{jk}$ are factored products of the theta functions
 $H(a\eta)$ and $\Theta(a\eta).$ These two properties make
${\tilde R}^{(j)}(\mu,\nu)$ much simpler for
$j=4,5$ than the original matrices $R^{(j)}(\mu,\nu).$
It is our belief that these ${\tilde R}^{(j)}(\mu,\nu)$
are the ``simplest possible''
similarity transformations of  $R^{(j)}(\mu,\nu)$ and that this is the
form which should be generalized to arbitrary $j.$

\section{Comparison with Sklyanin}

An alternative approach to the fusion matrices has been given by Sklyanin
\cite{sk1},\cite{sk2} who,    instead of $R^{(j)}(\mu,\nu)$
considers the four matrices independent of the spectral variable $v$  
$S^{(j)}_k$ with $0\leq k \leq 3$ which
are defined by
\begin{eqnarray}
(a(v+(j-2)\eta))+b(v+(j-2)\eta))S_0^{(j)}&&=R^{(j)}(+,+)(v)+R^{(j)}(-,-)(v)\\
(a(v+(j-2)\eta))-b(v+(j-2)\eta))S_3^{(j)}&&=R^{(j)}(+,+)(v)-R^{(j)}(-,-)(v)\\
(c(v+(j-2)\eta))+d(v+(j-2)\eta))S_1^{(j)}&&=R^{(j)}(+,-)(v)+R^{(j)}(-,+)(v)\\
(c(v+(j-2)\eta))-d(v+(j-2)\eta))S_2^{(j)}&&=R^{(j)}(+,-)(v)-R^{(j)}(-,+)(v)
\end{eqnarray}
where $a(v),~b(v),~c(v)$ and $d(v)$ are defined by (\ref{bw}).
%(Note that our $v+(j-2)\eta$ is the variable $u$ of Sklyanin
%\cite{sk1},\cite{sk2}).

Sklyanin shows that in order for $R^{(j)}(\mu,\nu)(v)$ to satisfy the
Yang Baxter equation that the $S_k^{(j)}$ must be a representation of
the algebra
\begin{eqnarray}
S_\alpha S_0-S_0S_{\alpha}&&=
J_{\beta,\gamma}(S_{\beta}S_{\gamma}+S_{\gamma}S_{\beta})\nonumber\\
S_{\alpha}S_{\beta}-S_{\beta}S_{\alpha}&&=
(S_0S_{\gamma}+S_{\gamma}S_0)\label{skalg}
\end{eqnarray}
where $\alpha, \beta, \gamma$ are any cyclic permutation of $1,2,3$ and the
structure constants $J_{\alpha,\beta}$ satisfy
\begin{equation}
J_{12}+J_{23}+J_{31}+J_{12}J_{23}J_{31}=0
\end{equation}
and are explicitly computed in terms of theta functions. Sklyanin
demonstrated that the finite dimensional representations of this
algebra may be constructed from the space of theta functions with zero
characteristics of order $j$ as defined in \cite{w}.

We have verified that the matrices $S^{(j)}_k$ constructed from our
matrices ${\tilde R}^{(j)}(\mu,\nu)(v)$ do in fact satisfy Sklyanin's 
algebra although we are not aware of any direct proof that 
the fusion construction and the algebra (\ref{skalg}) are equivalent.
The relation of the functional equation (\ref{fun5}) to Sklyanin's
algebra also remains to be studied.

\section{The functional equation for $Q$ at $L=2$}

We conclude by proving relation (\ref{closing}) for L=2.
\begin{equation}
T^{(2)}(v-K)=e^{-iN\pi v/2K}Q_R(v)M^{-1}Q_L(v-iK')S.
\label{closingL2}
\end{equation}
We proceed in two steps by first explicitly computing the matrix $M$
and then proving that (\ref{closingL2}) holds.

\subsection{Computation of $M$}

The conjecture for the functional equation for $Q$ in (\ref{con}) is
not complete because we have not given an explicit form for the
normalizing matrix $A.$ Therefore because $M$ is computed from $A$ by
(\ref{mqdef}) our first task is to find a form for $M$ 
which is consistent with
the conjecture. 
For $L=2$ we do this by setting
$v=3K/2$ in (\ref{con3}) to find
\begin{equation}
M = \frac{\exp(-\frac{i3\pi N}{4})}{h^{N}(K)}Q_{L}(3K/2-iK')Q_R(K/2)
\label{M}
\end{equation}
The matrices $Q_R$ and $Q_L$ are defined by (\ref{qrsr}) and (\ref{qlsl})
where the matrices $S_R$ and $S_L$ contain the arbitrary parameters
$\tau_{\alpha,m}$ which are restricted only by the requirement that the
resulting matrices be nonsingular. We find it convenient to make the choice
\begin{equation}
\tau_{\gamma,0} = \delta_{\gamma,-1}, \hspace{0.2 in}
\tau_{\gamma,1} = \delta_{\gamma,-1}, \hspace{0.2 in}
\tau_{\gamma,-1} = \delta_{\gamma,1}, \hspace{0.2 in}
\tau_{\gamma,2} = \delta_{\gamma,1} \hspace{0.2 in}
\label{tau}
\end{equation}
and thus we have explicitly from (\ref{sr}) and (\ref{sl}) with $L=2$  
\begin{equation}
S_R(+,\beta)(v)=\left( \begin{array}{l l}
H(v+K)\delta_{\beta,-1} & H(v)\delta_{\beta,1}\\
-H(v)\delta_{\beta,-1} & -H(v+K)\delta_{\beta,1}\\
\end{array}\right),
\hspace{0.1 in}
S_R(-,\beta)(v)=\left( \begin{array}{l l}
\Theta(v+K)\delta_{\beta,-1} & \Theta(v)\delta_{\beta,1}\\
\Theta(v)\delta_{\beta,-1} & \Theta(v+K)\delta_{\beta,1}\\
\end{array}\right),
\end{equation}
\begin{equation}
S_L(\alpha,+)(v)=\left( \begin{array}{l l}
H(v-K)\delta_{\alpha,-1} & H(v)\delta_{\alpha,1}\\
-H(v)\delta_{\alpha,-1} & -H(v-K)\delta_{\alpha,1}\\
\label{sr2}
\end{array}\right),
\hspace{0.1 in}
S_L(\alpha,-)(v)=\left( \begin{array}{l l}
\Theta(v-K)\delta_{\alpha,-1} & \Theta(v)\delta_{\alpha,1}\\
\Theta(v)\delta_{\alpha,-1} & \Theta(v-K)\delta_{\alpha,1}\\
\end{array}\right)
\label{sl2}
\end{equation}
Thus we find
\begin{equation}
M_{\alpha\beta}= \frac{\exp(-\frac{i3\pi N}{4})}{h^{N}(K)}
{\rm Tr}m(\alpha_1, \beta_1)
m(\alpha_2, \beta_2)\cdots m(\alpha_N, \beta_N)
\label{M2}
\end{equation}
where 
\begin{eqnarray}
m(\alpha,\beta)&& =
 S_L(\alpha,+)(3K/2-iK')S_R(+,\beta)(K/2)
+S_L(\alpha,-)(3K/2-iK')S_R(-,\beta)(K/2)\nonumber\\ 
&&= -2iq^{-1/4}\exp(\frac{i\pi}{4})H(K/2)\Theta(K/2) \hat{m}(\alpha,\beta)
\label{mh}
\end{eqnarray}
with
\begin{equation}
\hat{m}(\alpha,\beta) = 
\left( \begin{array}{l l l l}
\delta_{\alpha,-1}\delta_{\beta,-1} & \delta_{\alpha,-1}\delta_{\beta,1} &
i\delta_{\alpha,1}\delta_{\beta,-1} &i \delta_{\alpha,1}\delta_{\beta,1} \\
0 & 0 & 0 & 0\\
0 & 0 & 0 & 0\\
i\delta_{\alpha,-1}\delta_{\beta,-1} & i\delta_{\alpha,-1}\delta_{\beta,1} &
\delta_{\alpha,1}\delta_{\beta,-1} & \delta_{\alpha,1}\delta_{\beta,1} \\
\end{array}\right)
\end{equation}
Only the first and the last column of $\hat{m}$ contribute
to trace in (\ref{M2}) and thus
\begin{equation}
M_{\alpha,\beta} =\left(-2{H(K/2)\Theta(K/2)\over q^{1/4}h(K)}\right)^N 
{\rm Tr}(\rho(\alpha_1,\beta_1)\cdots\rho(\alpha_N,\beta_N))
\end{equation}
where
\begin{equation}
\rho_{\alpha\beta}=
 \left( \begin{array}{l l l l}
\delta_{\alpha,-1}\delta_{\beta,-1} &  i\delta_{\alpha,1}\delta_{\beta,1} \\
i\delta_{\alpha,-1}\delta_{\beta,-1} &  \delta_{\alpha,1}\delta_{\beta,1} \\
\end{array}\right)
\label{m0}
\end{equation}
and 
\begin{equation} 
M^{-1}_{\alpha,\beta} =\left(-2{H(K/2)\Theta(K/2)\over q^{1/4}h(K)}\right)^{-N}
 {\rm Tr}(\rho(\alpha_1,\beta_1)\cdots\rho(\alpha_N,\beta_N))
\label{M_inv}
\end{equation}

\subsection{Computation of $e^{-iN\pi v/2K}Q_R(v)M^{-1}Q_L(v-iK')S$}

We now use (\ref{pqrl}),(\ref{sr2}), (\ref{sl2}) 
and (\ref{M_inv}) in the right
hand side of (\ref{closingL2}) to find 
\begin{equation}
e^{iN\pi v/2K}Q_R(v)M^{-1}Q_L(v-iK')S|_{\alpha,\beta}=
\left(-{h(K)\over 2H(K/2)\Theta(K/2)}\right)^N
{\rm Tr}X(\alpha_1,\beta_1)\cdots X(\alpha_N,\beta_N)
\label{help10}
\end{equation}
where
\begin{equation}
X_{\alpha,\beta} = 
e^{-i\pi v/2K}q^{1/4}\sum_{\gamma,\lambda}
S_{R}(\alpha, \gamma)(v)\rho(\gamma, \lambda)
S_L(\lambda, \beta)(v+2K-iK')
\label{xdef}
\end{equation}
which, using the notation,
\begin{equation}
H_{K} = H(v+K), \hspace{0.3 in} \Theta_{K} = \Theta(v+K),
\hspace{0.3 in} H = H(v), \hspace{0.3 in} \Theta = \Theta(v)
\end{equation}
is explicitly written as
\begin{equation}
X(+,+) = \sum_{\gamma,\lambda}
\left( \begin{array}{l l}
H_{K}\delta_{\gamma, -1} & H \delta_{\gamma, 1}\\
-H\delta_{\gamma, -1} & -H_{K} \delta_{\gamma, 1}\\
\end{array}\right) \otimes
\left( \begin{array}{l l}
\delta_{\gamma, -1}\delta_{\lambda, -1} 
& i \delta_{\gamma, 1}\delta_{\lambda, 1}\\
i \delta_{\gamma, -1}\delta_{\lambda, -1}
& \delta_{\gamma, 1}\delta_{\lambda, 1}\\
\end{array}\right) \otimes
\left( \begin{array}{l l}
\Theta_{K}\delta_{\lambda, -1} & i \Theta \delta_{\lambda, 1}\\
-i \Theta\delta_{\gamma, -1} & -\Theta_{K} \delta_{\gamma, 1}\\
\end{array}\right) 
\end{equation}
\begin{equation}
X(-,-) = \sum_{\gamma,\lambda}
\left( \begin{array}{l l}
\Theta_{K}\delta_{\gamma, -1} & \Theta \delta_{\gamma, 1}\\
\Theta\delta_{\gamma, -1} & \Theta_{K} \delta_{\gamma, 1}\\
\end{array}\right) \otimes
\left( \begin{array}{l l}
\delta_{\gamma, -1}\delta_{\lambda, -1} 
& i \delta_{\gamma, 1}\delta_{\lambda, 1}\\
i \delta_{\gamma, -1}\delta_{\lambda, -1}
&  \delta_{\gamma, 1}\delta_{\lambda, 1}\\
\end{array}\right) \otimes
\left( \begin{array}{l l}
H_{K}\delta_{\lambda, -1} & -i H \delta_{\lambda, 1}\\
-i H\delta_{\gamma, -1} &  H_{K} \delta_{\gamma, 1}\\
\end{array}\right) 
\end{equation}
\begin{equation}
X(+,-) = \sum_{\gamma,\lambda}
\left( \begin{array}{l l}
H_{K}\delta_{\gamma, -1} & H \delta_{\gamma, 1}\\
-H\delta_{\gamma, -1} & -H_{K} \delta_{\gamma, 1}\\
\end{array}\right) \otimes
\left( \begin{array}{l l}
\delta_{\gamma, -1}\delta_{\lambda, -1} 
& i \delta_{\gamma, 1}\delta_{\lambda, 1}\\
i \delta_{\gamma, -1}\delta_{\lambda, -1} 
&  \delta_{\gamma, 1}\delta_{\lambda, 1}\\
\end{array}\right) \otimes
\left( \begin{array}{l l}
H_{K}\delta_{\lambda, -1} & -i H \delta_{\lambda, 1}\\
-i H\delta_{\gamma, -1} &  H_{K} \delta_{\gamma, 1}\\
\end{array}\right) 
\end{equation}
\begin{equation}
X(-,+) =\sum_{\gamma,\lambda}
\left( \begin{array}{l l}
\Theta_{K}\delta_{\gamma, -1} & \Theta \delta_{\gamma, 1}\\
\Theta\delta_{\gamma, -1} & \Theta_{K} \delta_{\gamma, 1}\\
\end{array}\right) \otimes
\left( \begin{array}{l l}
\delta_{\gamma, -1}\delta_{\lambda -1} 
& i \delta_{\gamma, 1}\delta_{\lambda 1}\\
i \delta_{\gamma, -1}\delta_{\lambda, -1}
&  \delta_{\gamma, 1}\delta_{\lambda, 1}\\
\end{array}\right) \otimes
\left( \begin{array}{l l}
\Theta_{K}\delta_{\lambda, -1} & i \Theta \delta_{\lambda, 1}\\
-i \Theta\delta_{\gamma, -1} &  -\Theta_{K} \delta_{\gamma, 1}\\
\end{array}\right) 
\end{equation}
We note that
\begin{equation}
X(+,+)= 
\left( \begin{array}{r r r r r r r r}
H_{K}\Theta_{K}& 0 & 0 & 0 & 0 & 0 & 0 & -H \Theta\\
-iH_{K}\Theta&  0 & 0 & 0 & 0 & 0 & 0 & -iH \Theta_{K}\\
iH_{K}\Theta_{K}&  0 & 0 & 0 & 0 & 0 & 0 & iH \Theta\\
H_{K}\Theta&  0 & 0 & 0 & 0 & 0 & 0 & -H \Theta_{K}\\
-H\Theta_{K}& 0 & 0 & 0 & 0 & 0 & 0 & H_{K} \Theta\\
iH\Theta&  0 & 0 & 0 & 0 & 0 & 0 & iH_{K} \Theta_{K}\\
-iH\Theta_{K}&  0 & 0 & 0 & 0 & 0 & 0 & -iH_{K} \Theta\\
-H\Theta&  0 & 0 & 0 & 0 & 0 & 0 & H_{K} \Theta_{K}\\
\end{array}\right)
\end{equation}
The rank of $X(+,+)$ is two and thus  when used in the trace in 
(\ref{help10}) it may be replaced by
\begin{equation}
Z(+,+) = 
\left( \begin{array}{r r}
H_{K}\Theta_{K} & -H \Theta\\
-H\Theta & H_{K} \Theta_{K}\\
\end{array}\right)
\end{equation}
Similarly $X(+,-),~X(-,+),$ and $X(-,-)$ may be replaced by
\begin{equation}
Z(+,-) = 
\left( \begin{array}{r r}
H_{K}^2 & H^2\\
-H^2 & -H_{K}^2 \\
\end{array}\right)
\end{equation}
\begin{equation}
Z(-,+) = 
\left( \begin{array}{r r}
\Theta_{K}^2 & -\Theta^2\\
\Theta^2 & -\Theta_{K}^2 \\
\end{array}\right)
\end{equation}
\begin{equation}
Z(-,-) = 
\left( \begin{array}{r r}
\Theta_{K}H_{K} & \Theta H\\
\Theta H & \Theta_{K}H_{K} \\
\end{array}\right)
\end{equation}
and therefore we have
\begin{equation}
e^{iN\pi v/2K}Q_R(v)M^{-1}Q_L(v-iK')S|_{\alpha,\beta}=
\left(-{h(K)\over 2 H(K/2)\Theta(K/2)}\right)^N
{\rm Tr}Z(\alpha_1,\beta_1)\cdots Z(\alpha_N,\beta_N)
\label{help20}
\end{equation}

\subsection{Proof of (\ref{closingL2})}

We recall from (\ref{trj})
and (\ref{bwr2}) that
\begin{equation}
T^{(2)}|_{\alpha,\beta}(v-K)={\rm Tr}R^{(2)}(\alpha_1,\beta_1)(v-K)\cdots
R^{(2)}(\alpha_N,\beta_N)(v-K)
\label{help21}
\end{equation}
with 
\begin{eqnarray}
&&R^{(2)}(+,+)(v-K) = 
\left( \begin{array}{c c}
a(v-K) & 0 \\
0 & b(v-K) \\
\end{array}\right),
%\hspace{0.3 in}
R^{(2)}(+,-)(v-K) = 
\left( \begin{array}{c c}
0 & d(v-K) \\
c(v-K) & 0 \\
\end{array}\right),\nonumber\\
&&R^{(2)}(-,+)(v-K) = 
\left( \begin{array}{c c}
0 & c(v-K) \\
d(v-K) & 0 \\
\end{array}\right),
%\hspace{0.3 in}
R^{(2)}(-,-)(v-K) = 
\left( \begin{array}{c c}
b(v-K) & 0 \\
0 & a(v-K) \\
\end{array}\right)
\end{eqnarray}

We thus complete the proof of (\ref{closingL2}) by noting that there
is a similarity transformation by a matrix $G$
which maps the matrices $Z(\alpha, \beta)$ on the matrices 
$R^{(2)}(\alpha, \beta).$
Specifically
\begin{equation}
GZ(+,+)G^{-1} = -\frac{H(K)}{H(K/2)\Theta(K/2)}
\left( \begin{array}{c c}
a(v-K) & 0 \\
0 & b(v-K) \\
\end{array}\right)
\label{zrsim1}
\end{equation}
\begin{equation}
GZ(+,-)G^{-1} = -\frac{H(K)}{H(K/2)\Theta(K/2)}
\left( \begin{array}{c c}
0 & d(v-K) \\
c(v-k) & 0 \\
\end{array}\right)
\label{zrsim2}
\end{equation}
\begin{equation}
GZ(-,+)G^{-1} = -\frac{H(K)}{H(K/2)\Theta(K/2)}
\left( \begin{array}{c c}
0 & c(v-K) \\
d(v-k) & 0 \\
\end{array}\right)
\label{zrsim3}
\end{equation}
\begin{equation}
GZ(-,-)G^{-1} = -\frac{H(K)}{H(K/2)\Theta(K/2)}
\left( \begin{array}{c c}
b(v-K) & 0 \\
0 & a(v-K) \\
\end{array}\right)
\label{zrsim4}
\end{equation}
where
\begin{equation}
G = 
\left( \begin{array}{r r}
a_{11} & a_{11} \\
-a_{22} & a_{22} \\
\end{array}\right)
\hspace{0.3 in}
G^{-1} = 
\left( \begin{array}{r r}
a_{22} & -a_{11} \\
a_{22} & a_{11} \\
\end{array}\right)
\end{equation}
with
\begin{equation}
a_{11} = \frac{1}{\sqrt2}\left(\frac{H(K/2)}{\Theta(K/2)}\right)^{1/2}
\hspace{0.3 in}
a_{22} = \frac{1}{\sqrt2}\left(\frac{\Theta(K/2)}{H(K/2)}\right)^{1/2}
\end{equation}
The verification of this similarity transformation is 
straightforward by use of identities such as 
\begin{eqnarray}
H(v)\Theta(v)+H(v+K)\Theta(v+K) 
&&= {H(K)\Theta(K)\over H(K/2)\Theta(K/2)}H(v+K/2)\Theta(v-K/2)\nonumber\\
&&=-{H(K)\over H(K/2)\Theta(K/2)}b(v-K),
\end{eqnarray}
\begin{eqnarray}
H(v+K)\Theta(v+K)-H(v)\Theta(v)&& = 
-{H(K)\Theta(K)\over H(K/2)\Theta(K/2)}H(v-K/2)\Theta(v+K/2)\nonumber\\
&&-{H(K)\over H(K/2)\Theta(K/2)}a(v-K)
\end{eqnarray}
and 
\begin{equation}
H^2(v+K)+H^2(v)={H^2(K)\over
\Theta^2(K/2)}\Theta(v+K/2)\Theta(v-K/2)
={H(K)\over \Theta^2(K/2)}c(v-K).
\end{equation}

Thus using (\ref{zrsim1})-(\ref{zrsim4}) and (\ref{help21}) in 
(\ref{help20}) and
recalling the definition (\ref{hvdfn}) of $h(v)$ we find
\begin{equation}
e^{iN\pi v/2K}Q_R(v)M^{-1}Q_L(v-iK')S|_{\alpha,\beta}=
\left(\Theta(0)H^2(K)\Theta(K)\over 2H^2(K/2)\Theta^2(K/2)\right)^N
T^{(2)}_{\alpha,\beta}(v-K)
\end{equation}

Thus if we finally use the identity
\begin{equation}
\frac{\Theta(0)H^{2}(K)\Theta(K)}{2H^{2}(K/2)\Theta^{2}(K/2)} = 1
\end{equation}
which follows from (\ref{ad2}) with $a=K,~v=K/2$ and $u=-K/2$
we see that (\ref{closingL2}) is proven.

% yyyyyyyyyyyyyyyyyyyyyyyyyyyyyyyyyyyyyyyyyyyyyyyyyyyyyyy
\appendix

\section{Theta functions}

The definition of Jacobi Theta functions of nome $q$ is
\begin{eqnarray}
H(v)&=&2\sum_{n=1}^\infty(-1)^{n-1}q^{(n-{1\over 2})^2}\sin[(2n-1)\pi
v/(2K)]\\
\Theta(v)&=&1+2\sum_{n=1}^{\infty}(-1)^nq^{n^2}\cos(nv\pi/K)\nonumber\\
&=&-iq^{1/4}e^{\pi i v/(2K)}H(v+iK')
\label{thetadf}
\end{eqnarray}
where $K$  and $K'$ are the standard elliptic integrals of the first kind
and 
\begin{equation}
q=e^{-\pi K'/ K}.
\end{equation}
These theta functions satisfy 
\begin{equation}
\Theta(v)=\Theta(-v),~~~H(v)=-H(-v)
\end{equation}
and the quasi periodicity 
relations (15.2.3) of ref. \cite{baxb}
\begin{eqnarray}
H(v+2K)&=&-H(v)\label{hper1}\\
H(v+2iK')&=&-q^{-1}e^{-\pi i v/K}H(v)\label{Hper}
\end{eqnarray}
and
\begin{eqnarray}
\Theta(v+2K)&=&\Theta(v)\\
\Theta(v+2iK')&=&-q^{-1}e^{-\pi i v/ K}\Theta(v).
\end{eqnarray}
From (\ref{thetadf}) we see that $\Theta(v)$ and $H(v)$ are not
independent but satisfy
(15.2.4) of ref.\cite{baxb}
\begin{eqnarray}
\Theta (v\pm iK')=\pm iq^{-1/4}e^{\mp{\pi i v\over 2K}}H(v)\nonumber \\
H(v\pm iK')=\pm iq^{-1/4}e^{\mp{\pi i v\over 2K}}\Theta(v).
\label{threl}
\end{eqnarray}

\centerline{\bf ACKNOWLEDGMENTS}

\vspace{.2in}

We are pleased to thank Prof. T. Deguchi, Prof. M. Jimbo, 
Prof. T. Miwa, Prof. R.I. Nepomechie, and 
Prof. N. Yu. Reshetikhin for fruitful discussions. 
We are particularly indebted to Prof. R.J. Baxter for information on
theta functions and for bringing the functional equation (\ref{fun5})
to our attention. This work 
is partially supported by NSF grant DMR0302758.

\end{document}